# Dynamic Programming based Time-Delay Estimation (TDE) Technique for Analysis of Time-varying Time-delay

Deepak K. Gupta[1], George R. McKee and Raymond J. Fonck
*Department of Engineering Physics, University of Wisconsin, Madison, Wisconsin 53706
USA*

## *Abstract*

A new time-delay estimation (TDE) technique based on dynamic programming is developed to measure the time-varying time-delay between two signals. The dynamic programming based TDE technique provides a frequency response 5 to 10 times better than previously known TDE techniques, namely those based on time-lag cross-correlation or wavelet analysis. Effects of frequency spectrum, signal-to-noise ratio and amplitude of time-delay on response of the TDE technique (represented as Transfer Function) are studied using simulated data signals. The transfer function for the technique decreases with increase in noise in signal; however it is independent of signal spectrum shape. The dynamic programming based TDE technique is applied to the Beam-Emission-Spectroscopy (BES) diagnostic data to measure poloidal velocity fluctuations, which led to the observation of theoretically predicted zonal flows in high-temperature tokamak plasmas.

---

[1]   Present address: Tri Alpha Energy, Rancho Santa Margarita, California 92688 USA
Email: dpk011@yahoo.com

## *Introduction:*

Time-delay estimation (TDE) techniques have been used in various branches of physical science and technology. Traditionally these are used in the area of signal processing for communication and controlled processes [1]. However, in the last decade such techniques have also been adopted by plasma turbulence physicists to detect and explore new physical phenomena. In plasma physics, TDE techniques provide a unique tool to measure turbulence flow field. Investigation of the turbulence flow field is instrumental not only to understand turbulence in magnetically confined plasmas but also to experimentally verify theoretically predicted phenomenon like zonal flows [2], which are believed to play vital role in suppression of turbulence in magnetically confined plasmas.

Since last twenty years, Beam Emission Spectroscopy (BES) diagnostic in fusion plasmas has not only provided the measurements and understanding of plasma turbulence [3, 4] but has also served as a tool to measure and study the local magnetic field and beam-plasma interactions [5]. Time-delay estimation (TDE) techniques, based on time-lag cross-correlation analysis and wavelet analysis, have been applied in past to Beam Emission Spectroscopy (BES) diagnostic data to measure the turbulence flow field [6-8]. In DIII-D tokamak, BES diagnostic measures spatially resolved density fluctuation in a two-dimension (radial-poloidal) plane. Velocity fluctuation of underlying density structure in the poloidal direction is directly related to the time-varying time-delay between poloidally separated channels. The application of a one-dimension TDE technique on this data provides a time-varying time-delay between two poloidally located channels, and hence a time dependent measurements of turbulence poloidal flows. The measurements of turbulence flow field using TDE techniques have allowed experimental confirmation of the presence of Geodesic Acoustic Modes (GAM) in high-temperature tokamak plasmas [8-11]. Recently, zero-mean-frequency (or, low-frequency) zonal flow has also been observed by applying the TDE techniques on enhanced BES diagnostic [12, 13]. The new TDE technique used in this work is based on the Dynamic Programming algorithm [14] and is presented in detail here.

The novel TDE technique described in the present article uses the Dynamic Programming algorithm, which is innovated from its use in fluid dynamics for the measurement of velocity field by searching for an optimal match for displacements allowed only in time [15]. This technique does not require any transformation of data into frequency or wavelet regime, and provides a much higher frequency response compared to previously known TDE techniques based on time-lag cross-correlation and wavelet analysis. Also, its response curve (represented as a transfer function) is independent of the shape of the carrier signal spectrum. For BES diagnostic application, spatially separated time-series signals of density fluctuations serve as carrier signals for the measurement of time-varying time-delay. Experiences with previous time-lag cross-correlation based and wavelet-based TDE methods [16] have shown that accuracy and response for estimating time-varying time-delay strongly depend upon 1) the frequency spectrum of carrier signals, 2) the signal-to-noise ratios (SNR) of carrier signals, and 3) the amplitude of time-varying time-delay. Current article describes the Dynamic Programming TDE technique and its implementation. The article also presents the detail characteristics of dynamic programming based TDE technique in terms of carrier spectrum, time-delay amplitude, and noise in carrier data.

## *Dynamic Programming TDE Technique and its Implementation:*

Dynamic programming can be described as an iterative optimization mathematical technique for solving a problem consisting of overlapping subproblems with an optimal substructure. The technique breaks the large problem down into incremental steps such that, at any stage, an optimal solution to the subproblem is known. In a bottom-up approach of dynamic programming algorithm, used in the current procedure, subproblems are solved multiple times in order to solve the complete problem. Here it is used to estimate the time-delay between two coherent, one-dimensional time signals obtained from density fluctuation measurements using BES diagnostic. In the present TDE technique, dynamic programming works as a vector matching algorithm to estimate temporal shift for best match between coherent parts of two signals. This time varying temporal shift directly corresponds to the time-varying time-delay and, gives velocity fluctuations between signals. A similar technique has been used in past for particle imaging velocimetry (PIV) in experimental fluid dynamics, where it is applied to two dimensional spatial data [15]. Here the technique is applied to temporal data to estimate a time-delay that itself fluctuates in time, *i.e.*, time-varying time-delay.

To calculate the time-delay between two coherent time signals (say, *U(i)* and *V(j)*), using dynamic programming algorithm based TDE technique, first a "local matching residual" function matrix, *d(i,j)*, is calculated by finding the absolute difference between the signals at all possible shifts, i.e., *d(i,j)=|U(i)-V(j)|*. A graphical representation of this matrix is shown in figure 1. The size of this matrix is proportional to the record length of the signals (i.e., *max(i) × max(j)* ). For two identical and non repeatable signals, which have zero time-delay in between them, the minimum values of the local matching residual, *d(i,j)* will lie in a straight line along the diagonal of this matrix (*i.e.*, *i=j*). For a non-zero constant time-delay ($k_\tau$), the minimum values of local matching residual will lie in a diagonally shifted straight line, and the shift will be proportional to the constant time-delay between signals (*i.e.*, $i=j\pm k_\tau$). However, for a time-varying time-delay, the minimum values in the matrix will not follow a straight line path. Moreover, depending upon the characteristics of the signals, there may be multiple paths of minimum local residuals. However, an optimum path will be the one along which the sum of all local matching residuals will be minimum. This is also true for the cases of constant time-delays (of $k_\tau$ including zero) where minima fall in a straight line. The dynamic programming algorithm provides an efficient way to find the optimal path, which has a minimum total of the local matching residual. The sum over the optimal path of the local matching residual function is defined as the "Accumulated Residual Function", *D(i,j)*, that follows the recursive equation:

$$D(i,j) = \min \begin{cases} D(i, j-1) + d(i, j-1) + d(i,j) \\ D(i-1, j-1) + 2\times[d(i-1, j-1) + d(i,j)] \\ D(i-1, j) + d(i-1, j) + d(i,j) \end{cases}.$$

The initial conditions used to recursively calculate the *D(i,j)* function are

$$D(i,j) = \begin{cases} 0, & \text{if } i+j = m \\ \infty, & \text{if } |i-j| > m \text{ or } (i+j) < m \end{cases},$$

where, $m$ is the maximum absolute displacement selected to search the matching path within the neighborhood of the *i=j* diagonal (see, Fig.1). A search of the optimal path is also subjected to appropriate continuity and boundary conditions. Above procedural steps can be iteratively repeated, even with interpolation, to achieve a temporal precision and fractional time-step resolution [15].

An example of this approach of dynamic programming based TDE analysis applied for a simple case of sinusoidal signal has been presented in reference [17], where a sinusoidal time-varying time-delay signal was imposed on a carrier sinusoidal signal. It has been shown that with delay frequency to carrier frequency ratio of 0.5, dynamic programming based TDE provides a near perfect extraction of input signal. The same article [17] also discussed the case of a sinusoidal time-delay imposed on a Gaussian carrier, and the spectral dispersion for various delay frequencies. In the present article the dynamic programming based TDE method is expanded much further, and its characteristics and capabilities are studied in detail.

For the present study of the dynamic programming based TDE technique, test signals are created that are closer to the real experimental situation, and relevant for application with BES diagnostics. Two pairs of simulated delay signals are created using the procedure shown by the flow diagram in figure 2. First, a carrier signal is selected. For most of the work presented in this article a shifted Gaussian is chosen, which closely matches a typical turbulence signal spectrum (Fig 3(a)). The carrier signal is created in frequency domain using a frequency spectrum of shifted Gaussian spectrum. A random phase is used such that it gives only real time series for the carrier signal in time domain. Two similar white noise signals are created, each with same amplitude and upper cutoff frequency at 0.98 times of Nyquist frequency ($f_{Ny}$), but these differ from each other by statistically different random phases. The carrier signal is added to these white noise signals to get a set of two simulated noisy carrier data signals. Delay signals of half the desired delay amplitude and opposite directions (i.e., delayed & forwarded) are imposed using sinc function [7] on each of these simulated carrier data signals, which have statistically different noises. This gives two pairs of simulated turbulence data signals with statistically different noise, where each pair of signals has same desired delay. As the imposition of time varying time-delay modifies the spectrum of the simulated data signal, it is found that the approach to impose the half delay amplitude in opposite directions on two signals instead of one full amplitude delay on one signal is more accurate in response and leads to spectrally identical signals. The TDE technique is applied to both sets of simulated turbulence data signals to estimate the imposed time-delay. The use of both sets helps to minimize the accumulating numerical noise, which is discussed in detail later. Imposed and estimated time-delays are compared and analyzed for different analysis parameters and experimental conditions.

Simulated turbulence data signals with a total data length of 262144 ($= 2^{18}$) points are used. For computational efficiency, data length is sequentially broken in 256 datasets of equal length. TDE is performed on each dataset of length 1024 points, and results are combined together in same order. To be compatible with the BES diagnostics turbulence signal, a sample-time, $\Delta t$, of one microsecond is used. Also, similar to actual BES turbulence signals, simulated data signals are filtered using an elliptical filter of order five with bandwidth of 10Hz to 400 kHz. For better accuracy and to minimize numerical rounding errors, data signals are normalized so that the amplitude ranges of both the signals are same and nominal. Unless specified otherwise, a wideband flat power spectrum time-delay signal with rms amplitude of $0.7\Delta t$ and zero mean is imposed on carrier signals, and used throughout this article. This also implies that the absolute amplitude of the delay signal, almost all of the time (i.e., 99.7%), is under $2.1\Delta t$ ( $=3\sigma$) by considering that it has a Normal distribution. For all wideband flat spectrum signals used in the current article, either for imposed time-delay signal, carrier signal or white noise, upper cutoff frequency is defined by applying a Butterworth filter of order ten. A typical example of this is an imposed time-delay signal with an upper cutoff frequency of 0.50 times the Nyquist frequency, ($f_{Ny}$), and is shown in figure 3(a).

An iterative process approach is employed to achieve a sub sample-time ( $<\Delta t$) resolution and accuracy of time-delay estimation. Sub sample-time resolution is also an essential requirement for the application of the present technique to BES diagnostic data for turbulence study. Current implementation of dynamic programming based TDE algorithm divides the sample-time by half in every iteration, i.e., ($\Delta t$ to $\Delta t/2$). After every TDE iteration, input simulated data time signals are corrected by the estimated time-delay and used as the new input data time signals for the next iteration. This process is repeated till a desired resolution and/or accuracy is achieved. Accuracy of the estimation is also improved by smoothing the data and limiting the estimation of time-delay to an experimentally feasible bandwidth. The work presented in this article uses only two iterations and provides an effective resolution of $0.25\Delta t$ or better due to smoothing at every iteration. In certain cases, an initial interpolation of the data signal, before applying the TDE algorithm, can further improve resolution and accuracy. The initial value of the maximum absolute displacement, $m$, is related to the maximum amplitude to be searched in the residual matrix of the signals. For the present case, it is selected to be $2\Delta t$. In every consecutive iteration, a new value of m is defined by $m=2(m-1)$. With the reduction of sample-time by half in each iteration, this effectively reduces the maximum search amplitude for every successive iteration.

### *Transfer Function:*

To quantify the response of a TDE method, a transfer function is evaluated by comparing the input time-delays with the estimated output time-delays, at various frequencies. Mathematically, the transfer function is defined as

$$R(f) = \sqrt{\frac{\langle T_{o1}(f) \cdot T_{o2}(f) \rangle}{\langle T_i(f) \cdot T_i(f) \rangle}} ,$$

where, the denominator inside the square root is auto-power of the input time-delay signal, $T_i$, in frequency domain. The numerator inside the square root is a cross-power of the pair of output estimated time-delay signals, $T_{o1}$ and $T_{o2}$, in frequency domain. Both these estimated output signals are similar, except that each one has statistically different and uncorrelated noise. Estimated output time-delay signals may also contain a numerical noise due to computation and calculation for the TDE technique itself. However, use of cross-power for the calculation of transfer function minimizes its contribution. For the present work, this calculation also puts forward the requirement of having two similar sets of simulated input (carrier) data signal pairs with uncorrelated noise (Fig.2).

For application of TDE techniques, transfer functions help to characterize and compare analysis parameters and the TDE techniques themselves. Figure 3 compares the transfer functions for three TDE techniques for noise free carrier signals. For the dynamic programming TDE technique the noise free carrier signal is a frequency shifted Gaussian function with a peak at $0.115f_{Ny}$ and a half width of $0.436f_{Ny}$, as shown in figure 3(a). This is well represented by a typical density fluctuation signal of BES diagnostics, which is also shown in figure 3(a) and used as the carrier for the case of Time-lag cross-correlation and wavelet based TDE techniques [16]. The input delay signal is a wideband flat power spectrum signal with a rms amplitude of $0.7\Delta t$ and a upper cutoff frequency of $0.5f_{Ny}$.

The transfer function of the time-lag cross-correlation TDE method is close to unity for very low frequencies of time-delay, which represent the faithful extraction of time-delay at these low frequencies. However, at somewhat higher frequencies (i.e., $> 0.06f_{Ny}$), the transfer function for the time-lag cross-correlation TDE method falls sharply and reaches below 0.2. The wavelet based TDE has a comparatively moderate transfer function at lowest frequencies. However with an increase in frequency, the drop in transfer function is noticeably lower compared to the time-lag cross-correlation TDE method. The transfer function for the wavelet method stays above 0.2 even after the frequency of $0.1f_{Ny}$. This is a nearly two-fold extension in the frequency range of TDE capabilities with wavelet analysis [16].

The dynamic programming based TDE analysis not only gives a near perfect transfer function for lower and intermediate frequencies, but also extends the response to much higher frequency. The transfer function stays nearly the same up to $0.1f_{Ny}$ and decreases slowly after that to reach 0.2 only around $0.5f_{Ny}$. This is a major (5 to 10 times) improvement in frequency response capabilities of TDE methods, using dynamic programming algorithm. It should be noticed that around and above $0.5f_{Ny}$ the carrier signal power also becomes very low.

### *Effects of Time-Delay Amplitude:*

Figure 4(a) demonstrates the effect of rms amplitude of time-varying time-delay on the response of transfer function for dynamic programming TDE while keeping other signal and analysis parameters the same. For this analysis, flat wideband spectrum time-delay signals with a spectral cutoff of $0.5f_{Ny}$ and different rms amplitudes are used. Noise free simulated signals are created by imposing the time-delays on a noise free carrier function with a Gaussian spectrum of half width $0.436f_{Ny}$ and peak $0.115f_{Ny}$.

Figure 4(a) shows the deviation of the transfer function from the model transfer function (derived for an rms time-delay of $0.7\Delta t$) with the change in rms amplitude of the imposed time-delay. Input signals having a range of imposed time-delays from $0.4\Delta t$ to $3.0\Delta t$ are used. However, the same TDE analysis parameter values, those optimized and used for the model case of imposed time-delay of $0.7\Delta t$, are used. A deviation of the peak values of transfer functions from 1.0 (namely, at lower frequencies) implies a corresponding deviation from the faithful estimation of the time-delay. For a certain range of variation in the rms value of imposed time-delay, near $0.7\Delta t$, the peak transfer function values remain reasonably close to 1.0. The sensitivity of the change in transfer function is different for increasing and decreasing amounts of rms values of time-delay amplitude.

For a near doubling of the rms amplitude, only about 10% decrease in the peak value of the transfer function is observed. On the other hand, decreasing the rms amplitude by around half leads to a nearly 30% increase in the peak value of the transfer function. Largely, the amplitude of the transfer function changes with the change in time-delay amplitude, and the shape of the transfer function remains almost invariable. However, it is also noticeable that the model transfer function (with imposed delay of 0.7Δt and peak amplitude around 1.0) has the comparatively better frequency response. As the transfer function deviates from unity, the upper frequency cutoff of the transfer function decreases.

As discussed earlier, for the dynamic programming TDE technique, the maximum time displacement allowed in a "local matching residual" (displacement) matrix is defined by the value of the search range '$m$' (see Fig.1). For all the analyses shown in figure 4(a), $m$ is chosen to be 2Δt. The minimum time displacement in dynamic programming TDE is governed either by the time resolution of the dataset, or by the time step moved in each search step. For the present analysis, although the signal has a resolution of 1 microsecond (Δt=1μs), the time step is moved by 0.5Δt in each search step. This defines the resolution of the output data as 0.5Δt at each iteration. Also, the time series of the imposed flat wideband spectrum time-delay consist of changing amplitudes. For the time instances where peak (absolute) amplitude is higher than the $m$ value, the TDE technique is not able to sample the full amplitude, and a lower estimate of the transfer function is obtained. Conversely, for the time instances where the value of $m$ is much larger than the peak (absolute) amplitude of time-delay, an overestimate of time-delay is possible. A right choice for the values of $m$ (maximum time displacement) and the resolution of signal (or related minimum time displacement) with respect to time-delay values helps the dynamic programming based TDE analysis in achieving an appropriate transfer function with minimum overestimation or underestimation (i.e., peak values larger or smaller than 1.0).

To analyze these effects, transfer functions for signals with maximum and minimum imposed time-delay rms values in the analysis (i.e., 0.4Δt and 3.0Δt) are recalculated by optimizing the search range, *m*, and search step parameters. For the case of high time-delay of rms 3.0Δt, the maximum search range, *m*, is increased from 2Δt to 4Δt for accurately sampling the full amplitude of signals. For the case of low rms amplitude of 0.4Δt, an initial interpolation by two is performed on the input data signals, which reduces the initial Δt to 0.5μs from 1.0μs. Figure 4(b) and 4(c) show that these adjustments in the analysis parameters help in achieving a better estimation of the time-delay, and hence transfer functions with peak values closer to 1.0.

To a certain extent, the implementation of the dynamic programming algorithm with a sufficient number of iterations, along with consecutively reducing value of *m*, also helps minimize the under or over estimation problem of the transfer function. However, more iterations also add to unwanted numerical noise in the analysis. Under some situations, an overestimation may result from the use of unavoidable discrete time steps in the analysis. This overestimation, if present, may be more pronounced in the initial estimation iteration steps when search time step is larger to accurately estimate large time-delays, and the consecutive iterations with smaller Δt and *m* are not able to fully correct this overestimation.

### *Effect of Carrier Signal Spectrum Shape:*

In real situations, the carrier signal spectrum may change from system to system. It can even change in a single system based upon the system parameters. For BES diagnostic, the spectrum of density fluctuation may change noticeably with the change in plasma conditions, measurement location, etc. Hence, it is highly desirable to have a TDE method which can extract the time-varying time-delay from these signals that have different carrier shapes, with nearly similar accuracy and sensitivity. In addition to making the analysis simpler, the use of same analysis parameters helps in a comparison of the experimental procedures and results.

The transfer functions of representative carrier spectra used for comparison in the present TDE analysis are shown in figure 5(a). Two of the carrier spectrums are of Gaussian shapes, each with half widths of $0.436f_{Ny}$, having peaks at origin and $0.115f_{Ny}$ respectively. The other two carriers are flat wideband spectrums with high frequency cutoffs at $0.436f_{Ny}$ and $0.600f_{Ny}$. Noise free simulated signals are formed by imposing the time varying time-delay signal on each of these noise free carrier signals. The time varying time-delay signal having a flat wideband spectrum with rms amplitude $0.7\Delta t$ and cutoff at $0.5f_{Ny}$ is used (Fig. 3(a)). Identical analysis parameters are used to extract the imposed time-delay from these simulated signals using the dynamic programming based TDE technique. Figure 5(b) shows that the transfer functions for these signals, with different carrier signals, are very similar. Results do not show any significant impact on transfer function shape or amplitude due to modification in the carrier function, and the transfer function extends to nearly full spectral width of the imposed time-delay.

Figure 5(b) also shows a transfer function for the case where an imposed time-delay (flat wideband spectrum of amplitude $0.7\Delta t$) with $0.1f_{Ny}$ cutoff is used, instead of $0.5f_{Ny}$. This wideband flat spectrum delay with smaller cutoff frequency is imposed on a flat wideband carrier spectrum with the high frequency bandwidth of $0.436f_{Ny}$ (see Fig. 5(a)). The lower cutoff of the imposed time-delay reduces the spectral range for which the transfer function can be defined. However, the transfer function with lower cutoff still closely matches with the higher cutoff ($=0.5f_{Ny}$) transfer function up to the lower cutoff frequency of the imposed time-delay.

In this analysis with different carrier spectrums, most of the spectral range of imposed wideband flat spectrum time varying time-delay with $0.5f_{Ny}$ cutoff is within the non-zero spectral part of the carrier function. This is also true for most of the analysis presented in this article. Tests are also done for the special cases where the imposed time-delay with the frequency cutoff exceeded well above the carrier frequency cutoff. Such situations lead to a high spectral dispersion toward the low frequencies in the transfer function, resulting in much higher (greater than 1.0) and misleading values of the transfer function. A similar dispersive spectral behavior is noted when the cutoff frequency of a carrier spectrum approaches the Nyquist frequency (e.g., a wideband flat spectrum carrier with $0.8f_{Ny}$ cutoff), even when the spectral cutoff of the imposed delay spectrum is sufficiently low (e.g., wideband flat spectrum time- delay with $0.5f_{Ny}$ cutoff).

### *Effect of Noise in Carrier Signal:*

Noise in an experimental system is virtually unavoidable. In BES diagnostic system, a number of sources may contribute to the noise, and degrade signal-to-noise ratio (*SNR*). *SNR* can be improved by designing and implementing sophisticated and improved hardware [18, 19]. However, this not only heavily increases the cost of the diagnostics but is also limited by many practical constrains. Experiences with previous TDE techniques have shown that the capabilities of a TDE technique may also be limited by the presence of noise in the signal. Here we explore the effects of noise in carrier signals upon the transfer function for the dynamic programming based TDE technique.

In current analysis, the noise in simulated signals is represented by adding a flat wideband spectrum (white noise) of nearly full bandwidth, i.e., with a high frequency cutoff of $0.98f_{Ny}$. A noise factor, in the range of 0.1 to 2.0, is multiplied with the time amplitude of the noise signal to get different Signal-to-Noise Ratios (*SNR*). For the purpose of quantification and comparison with actual experimental data signals, a frequency dependent Signal-to-Noise Ratio, *SNR*, is defined as the ratio of the correlated component to the uncorrelated component of the (simulated) signal [16], i.e.,

$$SNR(f) = \frac{|N_c(f)|}{|N_u(f)|},$$

where, the correlated component, $N_c(f) = \sqrt{|\langle S_1(f) \cdot S_2(f) \rangle|}$, and, the uncorrelated component, $N_u(f) = \sqrt{|\langle S_1(f) \cdot S_1(f) \rangle| - |\langle S_1(f) \cdot S_2(f) \rangle|}$ either for signal $S_1$, or $N_u(f) = \sqrt{|\langle S_2(f) \cdot S_2(f) \rangle| - |\langle S_1(f) \cdot S_2(f) \rangle|}$ for signal $S_2$. The notation of angle brackets with a dot (< . >) implies auto-correlation or cross-correlation. The above definition of signal-to-noise ratio is common to simulated as well as experimental signals. For a simulated signal, figure 6 shows an example of a *SNR* plot along with its correlated and uncorrelated signals, having white noise with the noise factor 1.0. It can be noticed here that the *SNR* plot closely follows the correlated signal, mainly because the uncorrelated signal only consist of white noise in the simulated signal. However, this formulation of *SNR* should work with any kind of noise spectrum.

For studying the *SNR* effects on the dynamic programming based TDE technique, simulated data signals are formed with noise factors ranging from 0.1 to 2.0. The Signal-to-Noise Ratios (*SNR*) for these noise factors, for a Gaussian carrier signal with a peak at $0.115 f_{Ny}$ and spectral width $0.436 f_{Ny}$, are plotted in Figure 7(a). As intended, the amplitude of SNR decreases with an increase in the noise factor due to the corresponding increase in the amplitude of white noise.

Figure 7(b) shows the transfer functions for these simulated signals, calculated using the dynamic programming based TDE technique. The peak value for the transfer function is highest and closest to 1.0 for the lowest noise factor or highest *SNR*. The peak value of the transfer function decreases with an increase in noise factor; however the shape of the transfer function remains largely same for all values of the noise factor. A decrease in the frequency response of the transfer function can also be noticed with a decrease in its maximum amplitude.

The decrease in the amplitude of the transfer function with an increase in noise in signal imposes a limit on the *SNR* level for which a reasonable quality of time-delay estimation can be achieved by a TDE technique. For the specific example used in figure 7, the peak value of the transfer function falls below 0.2 for a noise factor of 1.0 and higher, which corresponds to maximum *SNR* values of nearly 2 or lower. The correlated and uncorrelated signals for a noise factor value of 1.0 are shown in figure 6.

### *Application to BES diagnostic data:*

The Dynamic Programming based Time-delay Estimation technique, described here, is applied to the experimental data collected using Beam Emission Spectroscopy (BES) in DIII-D tokamak for the detection and analysis of velocity fluctuation, which is related to plasma turbulence in high temperature plasmas. The technique has allowed the experimental observation of theoretically predicted coherent zonal flows in turbulent plasmas. In the past, time-lag cross-correlation based and wavelet based TDE techniques have been used with BES data to detect Geodesic Acoustic Modes (GAM), also known as the high frequency branch of zonal flows [8-11]. The Dynamic Programming based TDE technique is applied to similar experimental data and the results are compared with the results previously obtained and reported with time-lag cross-correlation based and wavelet based TDE techniques.

Figure 8(a) shows the cross-coherency of poloidal velocity fluctuation in the poloidal direction at gradually increasing distances (1.1cm to 4.4cm). This clearly shows a coherent poloidal flow fluctuation around 15kHz, which is a strong indication of the presence of coherent poloidal fluctuation and, hence, the presence of GAM. Figure 8(b) further substantiates this presence of GAM by showing the near zero phase shift in the poloidal direction over the full range of measurements. The radial phase shift by $180^\circ$ within a distance of nearly 2.5cm is an expected observation as per the properties of GAM. The results presented in figure 8 are obtained using the Dynamic Programming TDE technique and are fully matched with the results previously reported on GAM using other TDE techniques on BES data on DIII-D tokamak [8-11, 20]. In addition, the dynamic programming based TDE technique has further extended the analysis capabilities to study GAM and plasma turbulence. Figure 8(c) shows the change in GAM frequency during a single plasmas shot using the dynamic programming based TDE technique. The change in GAM frequency is due to a gradual change in plasma temperature during this plasma shot, which is confirmed by the (ion & electron) temperature measurements using CER (Charge Exchange Recombination) and ECE (Electron Cyclotron Emission) diagnostics. Further analysis of the velocity fluctuation data using the dynamic programming based TDE technique shows an interesting dynamic of the wide-band frequency spectrum of poloidal velocity fluctuation spectrum flowing in the poloidal direction. Such wide-band frequency spectrum of poloidal velocity fluctuation have also been noticed in past using other TDE techniques [16, 20].

Dynamic programming based TDE technique has also facilitated the observation of theoretically predicted zero-mean-frequency (or low-frequency) zonal flow [2]. Recently we have reported [12, 13] the observation of this zero-mean-frequency (low-frequency) zonal flow in a high temperature plasma of DIII-D tokamak, utilizing enhanced BES diagnostics along with Time-Delay-Estimation techniques. The newly installed enhanced BES diagnostics provide a signal-to-noise ratio (SNR) an order of magnitude higher than the previous system [19]. Both time-lag cross-correlation based and dynamic programming based TDE techniques are used in the estimation of poloidal velocity fluctuation for the detection of low-frequency zonal flow. Both TDE techniques provide equivalent results in the low frequency range of interest. At higher frequency range, the dynamic programming based TDE technique provides a unique insight into these measurements.

## *Conclusion:*

A new Time-Delay Estimation (TDE) technique using dynamic programming algorithm is developed. This technique provides a better response (transfer) function compared to previously known TDE techniques based on time-lag cross-correlation and wavelet-analysis. The dynamic programming based TDE technique not only provides higher sensitivity (or, higher values of the transfer function) at low and medium frequencies, but also exhibits a better response to frequencies many times higher than those possible with previously known TDE techniques.

The TDE technique based on dynamic programming is robust to moderate changes in average time-delay amplitudes in data, and does not require optimization of analysis parameters for small changes. For large changes in average time-delay amplitudes, optimization of analysis parameters related to search step and search range is simple to understand and apply.. The response of the transfer function of dynamic-programming based TDE technique is fairly independent of change in carrier function shape, which may occur, say, due to a change in experimental or parametric conditions during data collection. This allows incorporation of experimental variations without much effort or re-optimization of analysis parameters. It also makes the comparison of results under different conditions more reliable. However, similar to other TDE techniques [16], dynamic-programming based TDE technique is also sensitive to noise in carrier signal. A higher signal-to-noise ratio of carrier data leads to a better transfer function for the TDE technique.

The dynamic programming based TDE technique is applied to measure the poloidal velocity fluctuation in high temperature tokamak plasmas using beam-emission-spectroscopy diagnostics. The new technique is already providing useful and new information in the field of plasma turbulence in tokomaks. Extension of the TDE technique at much higher frequencies allows exploration and study of higher frequency phenomena, such as wide band poloidal velocity fluctuations in tokamak plasmas [16, 20]. The technique will continue to provide useful information in the field of plasma turbulence, and it can certainly be extended to other branches of physics and science.

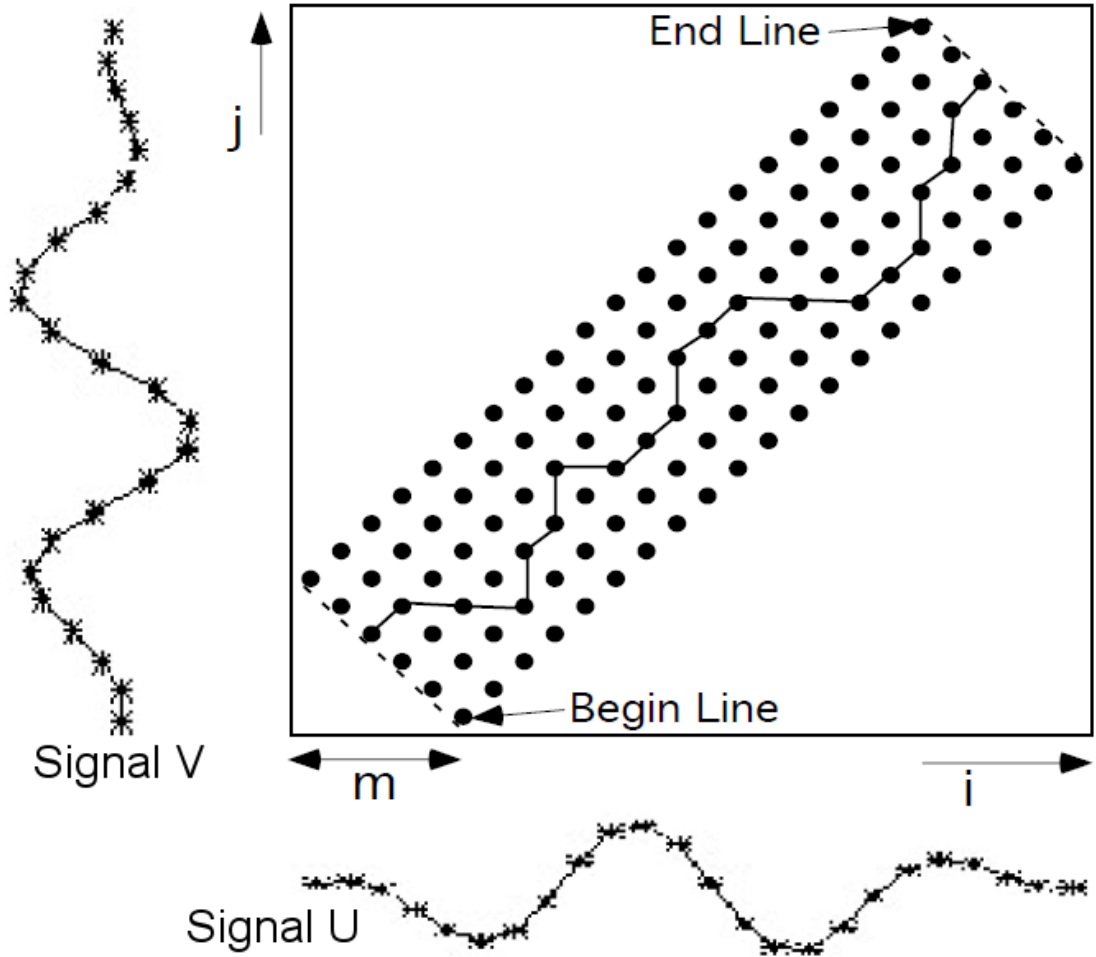

Figure 1: Illustration of Local Matching Residual matrix with a probable minimum path

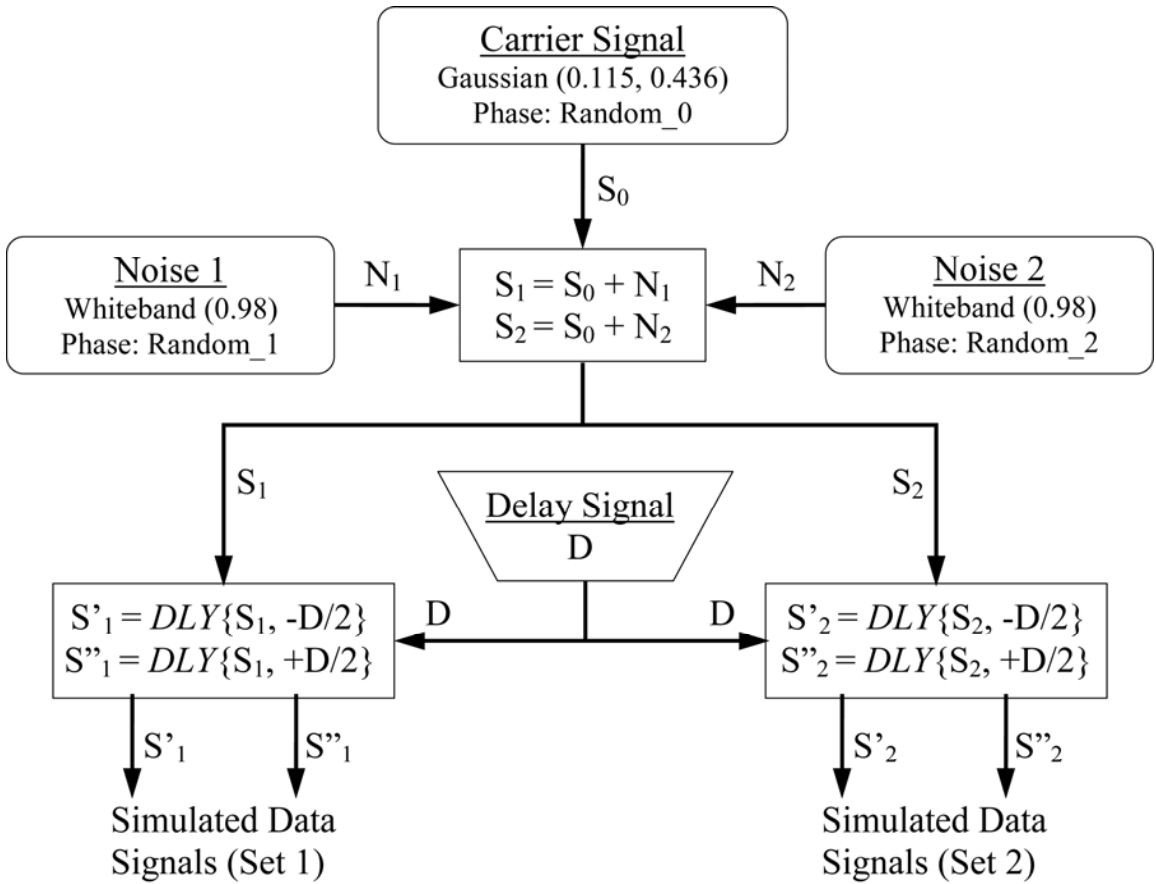

Figure 2: Flow diagram for creating two pair of simulated data signals, with same delay but statistically different noises.

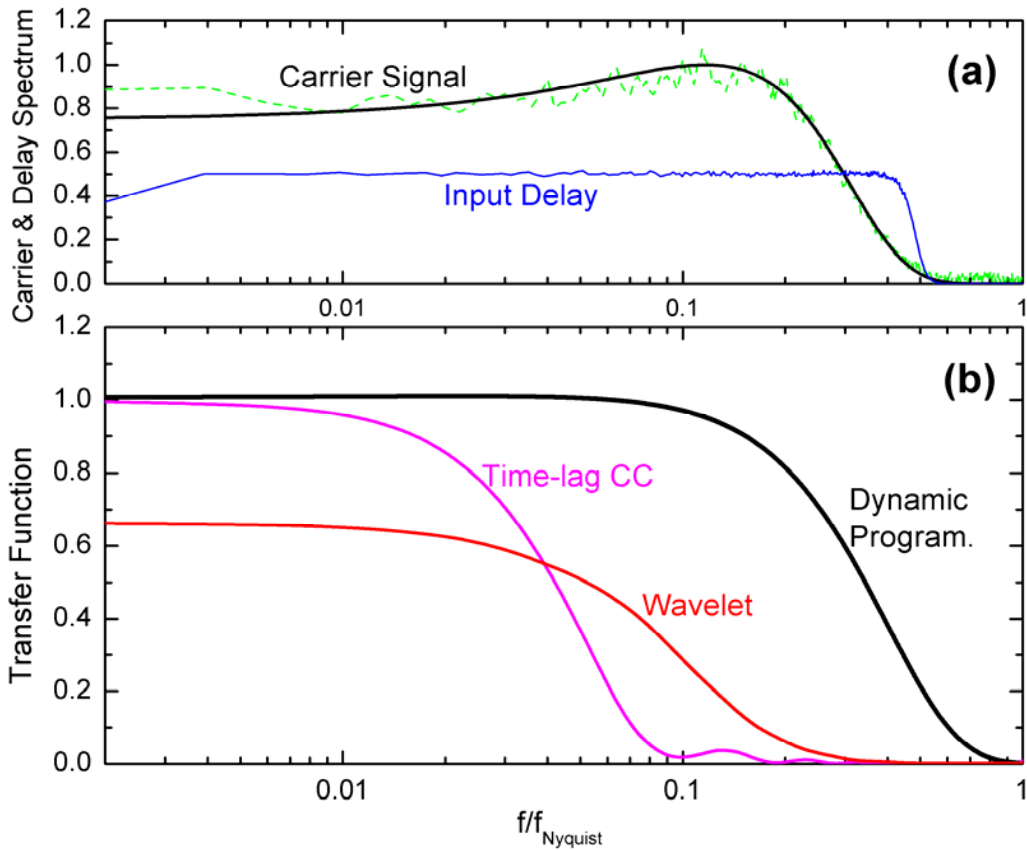

Figure 3: (a) Power spectrum of a typical turbulence signal from BES diagnostic [dotted line] along with the fitted Gaussian function used as the noise free carrier signal for dynamic programming based TDE technique. Also shown is the spectrum of input delay signal used in the calculations. (b) Comparison of different Time Delay Estimation (TDE) techniques.

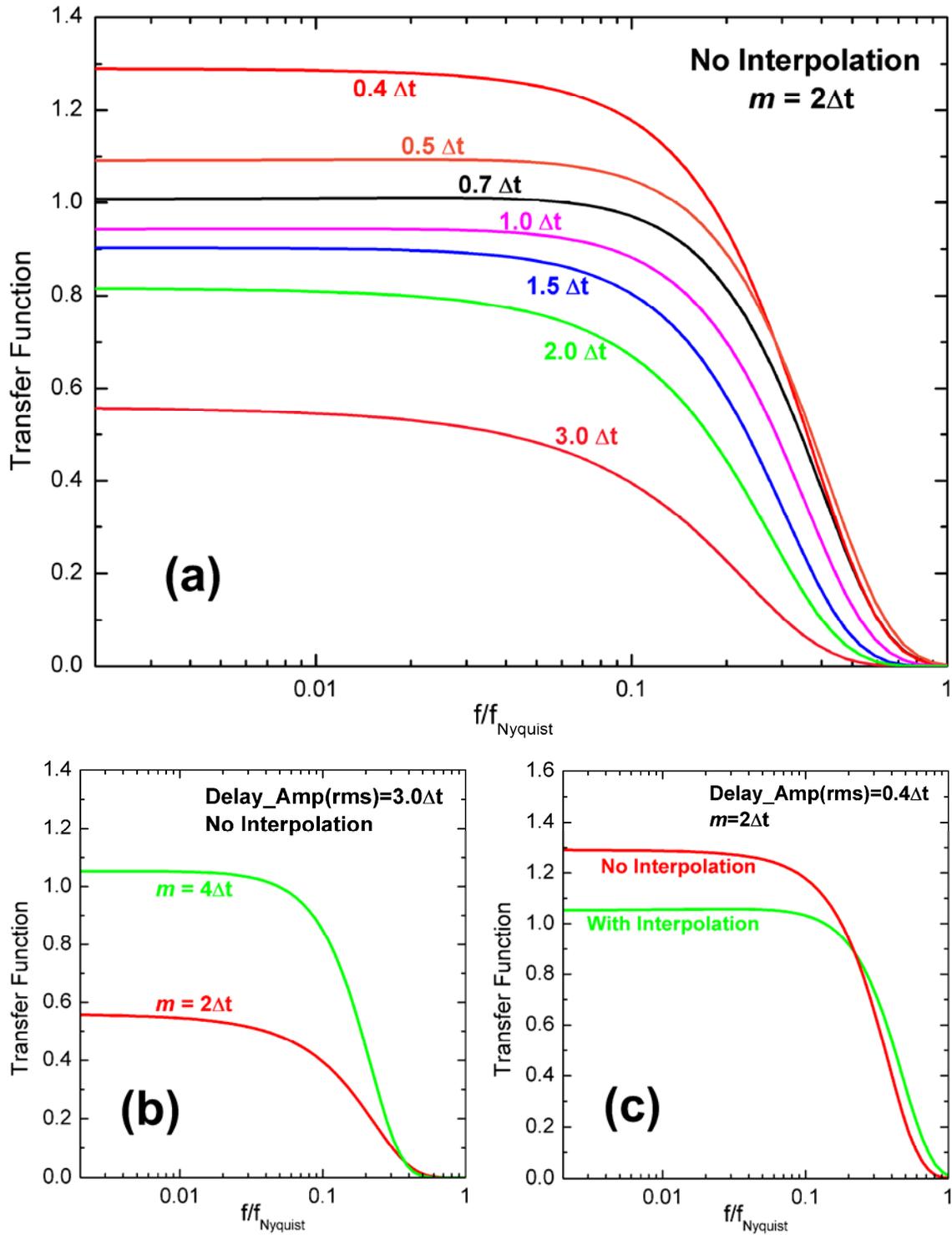

Figure 4: (a) Change in transfer function with the change in imposed time-delay amplitude (b) Improvement in transfer function with increase in *m* for high values of delay amplitude. (c) Improvement in transfer function with interpolation for small values of time-delay

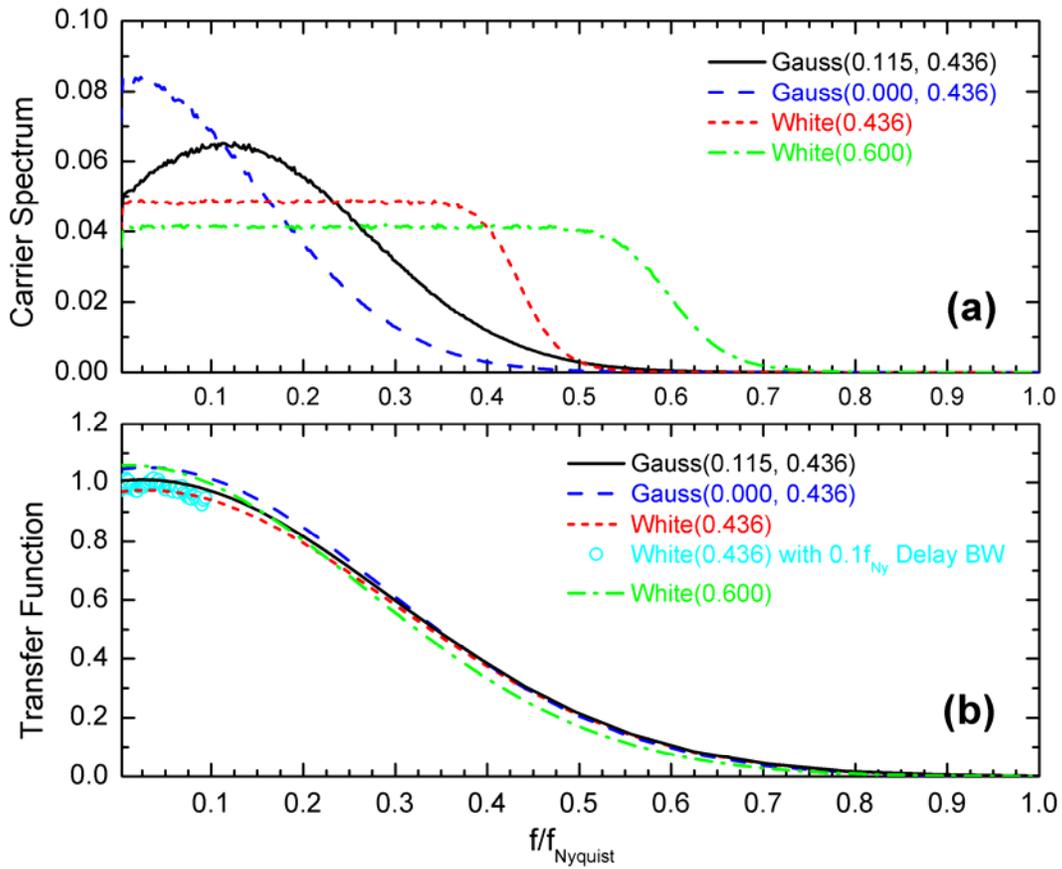

Figure 5: (a) Power Spectra of carrier signals used for comparing the effects of carrier spectra shape. (b) Transfer functions for the carriers shown in (a) along with one having a shorter bandwidth of $0.1 f_{Ny}$.

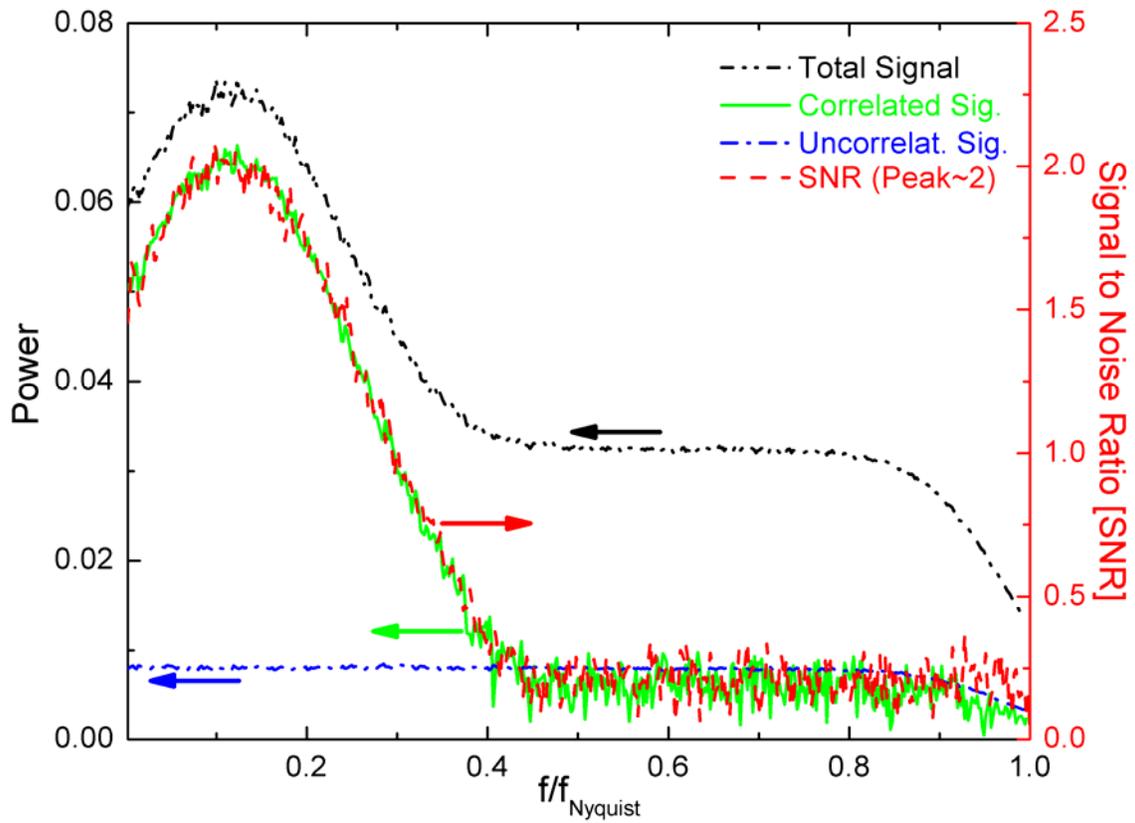

Figure 6: Correlated and uncorrelated components of the signal along with signal-to-noise ratio (*SNR*) plot, showing that the *SNR* closely follows the correlated signal.

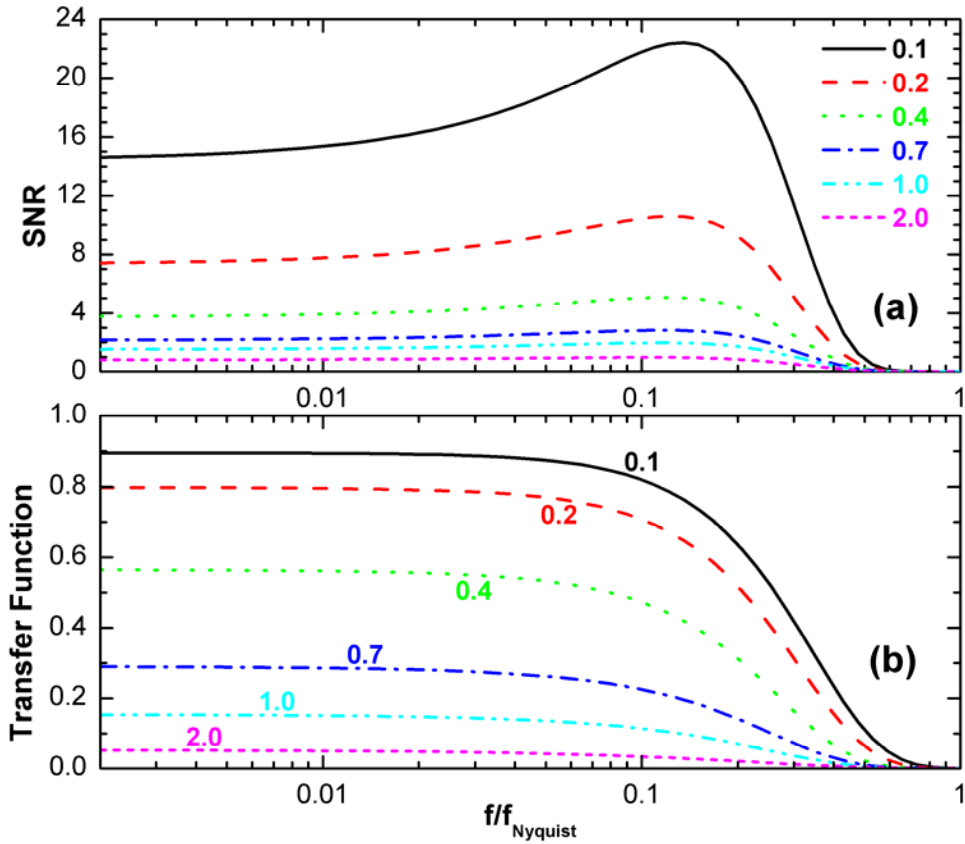

Figure 7: (a) Signal-to-Noise Ratio (*SNR*) plots for various Noise Factor multipliers of white noise added to the Gaussian carrier signal. (b) Transfer functions of dynamic programming TDE technique for *SNR* plots shown in (a).

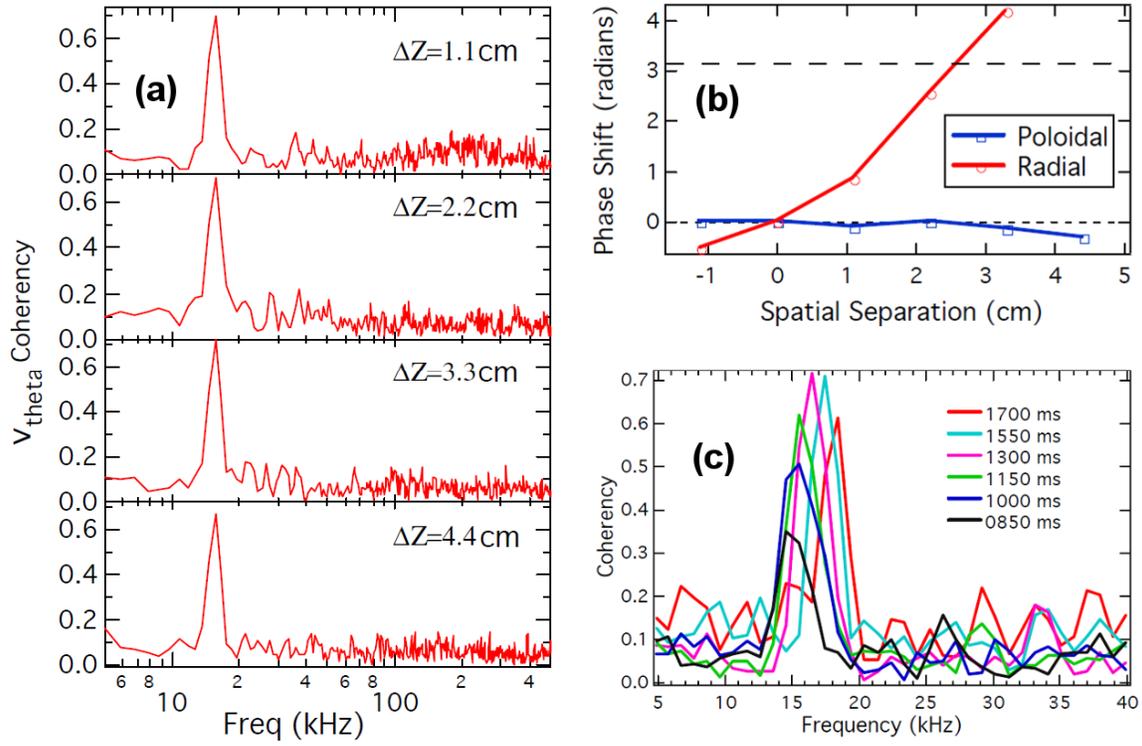

Figure 8: (a) Cross-coherency of poloidal velocity fluctuation at gradually increasing poloidal distances, showing Geodesic Acoustic Modes (GAM) as a coherent fluctuation around 15kHz. (b) Phase shifts of GAM in poloidal and radial directions. (c) Change in GAM frequency with the change in plasma temperature in a single plasma shot.

27